\documentclass[10pt]{article}
\usepackage{amsmath,amsfonts}
\usepackage{graphicx}
\usepackage[numbers]{natbib}
\usepackage{algorithm}
\usepackage[noend]{algpseudocode}
\usepackage{amssymb,graphicx}
\usepackage{mathtools}
\usepackage{mathptmx}
\usepackage[affil-it]{authblk}
\usepackage{amsthm}
\usepackage{amssymb}
\usepackage{mathrsfs}   % For \mathscr (script-like calligraphy)
\usepackage{bm} 
\usepackage{booktabs}
\usepackage{multirow,multicol}
\usepackage{hyperref}
\usepackage{adjustbox}
\usepackage{url}
\usepackage{physics} 
\usepackage[margin=1.10in]{geometry}

\title{\textbf{Deep Learning Approaches with Explainable AI for Differentiating Alzheimer Disease and Mild Cognitive Impairment}}

\author{
Fahad Mostafa\textsuperscript{1}, Kannon Hossain\textsuperscript{2}, Hafiz Khan \textsuperscript{3} \\

\textsuperscript{1}School of Mathematical and Natural Sciences, Arizona State University, USA \\
\textsuperscript{2}Department of Mathematics, Florida Gulf Coast University, USA \\
\textsuperscript{3}
Department of Public Health, Julia Jones Matthews School of Population and Public Health, Texas Tech University Health Sciences Center, USA \\

}
\date{}

\begin{document}

\maketitle

\begin{abstract}
Early and accurate diagnosis of Alzheimer Disease is critical for effective clinical intervention, particularly in distinguishing it from Mild Cognitive Impairment, a prodromal stage marked by subtle structural changes. In this study, we propose a hybrid deep learning ensemble framework for Alzheimer Disease classification using structural magnetic resonance imaging. Gray and white matter slices are used as inputs to three pretrained convolutional neural networks: ResNet50, NASNet, and MobileNet, each fine-tuned through an end-to-end process. To further enhance performance, we incorporate a stacked ensemble learning strategy with a meta-learner and weighted averaging to optimally combine the base models. Evaluated on the Alzheimer Disease Neuroimaging Initiative dataset, the proposed method achieves state-of-the-art accuracy of 99.21\% for Alzheimer Disease vs. Mild Cognitive Impairment and 91.02\% for Mild Cognitive Impairment vs. Normal Controls, outperforming conventional transfer learning and baseline ensemble methods. To improve interpretability in image-based diagnostics, we integrate Explainable AI techniques by Gradient-weighted Class Activation, which generates heatmaps and attribution maps that highlight critical regions in gray and white matter slices, revealing structural biomarkers that influence model decisions. These results highlight the framework’s potential for robust and scalable clinical decision support in neurodegenerative disease diagnostics.\\
{\textbf{keywords}}: Deep Learning, Alzheimer disease, Explainable Artificial Intelligence, Alzheimer Disease Neuroimaging Initiative
\end{abstract}

 \section{Introduction}
AD is a progressive neurodegenerative disorder and the most common cause of dementia worldwide, affecting millions of elderly individuals and placing immense socio-economic burdens on healthcare systems globally \cite{brookmeyer2007forecasting, prince2015world, winblad2016defeating}.  Alzheimer’s disease is a degenerative neurological condition that gradually impairs memory, reasoning, and the ability to carry out everyday activities \cite{knopman2021alzheimer}. It represents the leading cause of dementia, distinct from normal aging, and is linked to the accumulation of abnormal amyloid plaques and tau tangles in the brain, which harm and eventually destroy nerve cells. Common signs include forgetfulness, disorientation, difficulty with problem-solving or planning, and shifts in mood or personality, all of which intensify as the illness progresses \cite{heisterman2012cognitive}. Although there is currently no cure, available treatments may ease symptoms, and certain lifestyle adjustments can provide additional support. With the aging global population, early detection of AD has become a crucial area of research to enable timely interventions and slow disease progression \cite{day2024diagnosing}. Despite significant advances in biomarker discovery, the clinical diagnosis of AD remains challenging due to its complex pathology and overlapping symptoms with other cognitive disorders \cite{zhang2023artificial, frisoni2010clinical, jack2018nia}.

Descriptive epidemiological data on AD indicate that both prevalence and incidence rise steadily with advancing age, reaching their peak among the elderly, with women affected more often than men \cite{reitz2011epidemiology}. From a demographic perspective, AD poses a growing public health concern as populations continue to age, contributing heavily to illness and death rates \cite{henderson1986epidemiology}. Although regional variations are observed, the disorder affects communities worldwide, creating a considerable strain on caregivers as well as healthcare infrastructures \cite{zhang2021epidemiology}.

\begin{table}[h!]
\centering
\caption{List of abbreviations used in the manuscript.}
\begin{tabular}{ll}
\hline
\textbf{Abbreviation} & \textbf{Full Term} \\\hline
ML & Machine Learning\\
DL & Deep Learning\\
XAI &  Explainable Artificial Intelligence \\
AD   & Alzheimer Disease   \\
MCI    &  Mild Cognitive Impairment  \\
MRI   &  Magnetic Resonance Imaging  \\
ADNI   &  Alzheimer Disease Neuroimaging Initiative\\
NC  & Normal Controls \\
Grad - CAM & Gradient-weighted Class Activation \\
MMSE &  Mini-Mental State
Examination \\
CDR  & Clinical Dementia Rating  \\
CSF  &  Cerebrospinal Fluid  \\
PET  &  Positron Emission Tomography  \\
MoCA   &  Montreal Cognitive Assessment \\
LSME & Long Short-Term Memory  \\
LIME & Local Interpretable Model-Agnostic Explanations \\
LRP & Layer-wise Relevance Propagation \\
EADCD & Enhancing Automated Detection and Classification of Dementia \\
TIPAIT & Thinking Incapable People Using Artificial Intelligence Techniques \\
BGGO & Binary Greylag Goose Optimization \\
ISSA  & Improved Salp Swarm Technique \\
WNN & Wavelet Neural Network \\
ConvNets   &  Convolutional Neural Networks  \\
GM & Gray Matter\\
WM & White Matter\\
ROC & Receiver Operating Characteristic \\
AUC & Area Under the Curve\\
\hline
\end{tabular}
\label{tab:abbreviations}
\end{table}

Traditionally, AD diagnosis has relied on cognitive tests such as the  MMSE, CDR, and invasive tests like CSF biomarker assays and  PET scans \cite{morris1993clinical, folstein1975mini, mantzavinos2017biomarkers}. While accurate, these procedures are often expensive, invasive, and unavailable in primary care settings. As a result, MRI-based analysis has gained traction due to its non-invasive nature and ability to reveal structural changes in the brain, such as hippocampal atrophy, that are indicative of AD \cite{cuingnet2011automatic, liu2022generalizable, razavian2015population}. 

Alzheimer’s disease can be identified through a combination of clinical, cognitive, and biological assessments \cite{snyder2014assessing, corey1995diagnosis}. Physicians typically begin with a detailed medical and family history alongside evaluations of symptom progression, focusing on memory decline and difficulties in daily functioning. Cognitive screening tools, such as the MMSE or the MoCA \cite{larner2012screening}, are frequently employed to measure impairments in memory, language, and reasoning. Neurological and physical examinations help exclude other potential causes of cognitive decline, while neuroimaging techniques like MRI or CT scans can reveal brain atrophy and rule out alternative conditions \cite{small2008current}. Advanced imaging, such as PET scans, may further detect amyloid plaques or tau tangles characteristic of AD \cite{wolf2003critical, tigano2019neuroimaging}. In addition, biomarker analyses of cerebrospinal fluid or blood provide evidence of abnormal amyloid and tau protein levels, and in rare hereditary cases, genetic testing may be used to confirm early-onset disease. Together, these methods contribute to a comprehensive and reliable diagnosis of AD.

Over the past decade, DL has emerged as a transformative approach for medical image analysis. Convolutional neural networks, in particular, have demonstrated remarkable success in detecting patterns in MRI scans for AD classification \cite{ji2019early, suk2017deep, shi2017multimodal}. Compared to traditional ML methods, DL techniques eliminate the need for handcrafted features, instead learning hierarchical representations directly from the raw imaging data \cite{ litjens2017survey, thibeau2023interpretability}. These models can capture subtle structural changes associated with different stages of AD, including NC, MCI, and AD itself \cite{alsubaie2024alzheimer, bossa2023multidimensional, cheung2022deep}.

Despite these promising developments, several challenges persist. Many CNN-based studies have limited generalizability due to small sample sizes or single-source datasets. Furthermore, standalone models often suffer from reduced accuracy when applied to real-world clinical settings with diverse imaging protocols and patient demographics \cite{liu2022generalizable, mmadumbu2025early}. To address these limitations, ensemble methods that combine multiple models have been explored, leveraging model diversity to improve robustness and performance \cite{ji2019early, mmadumbu2025early}.
\begin{figure}[htbp]
	\centering
	\includegraphics[width=1.0\textwidth]{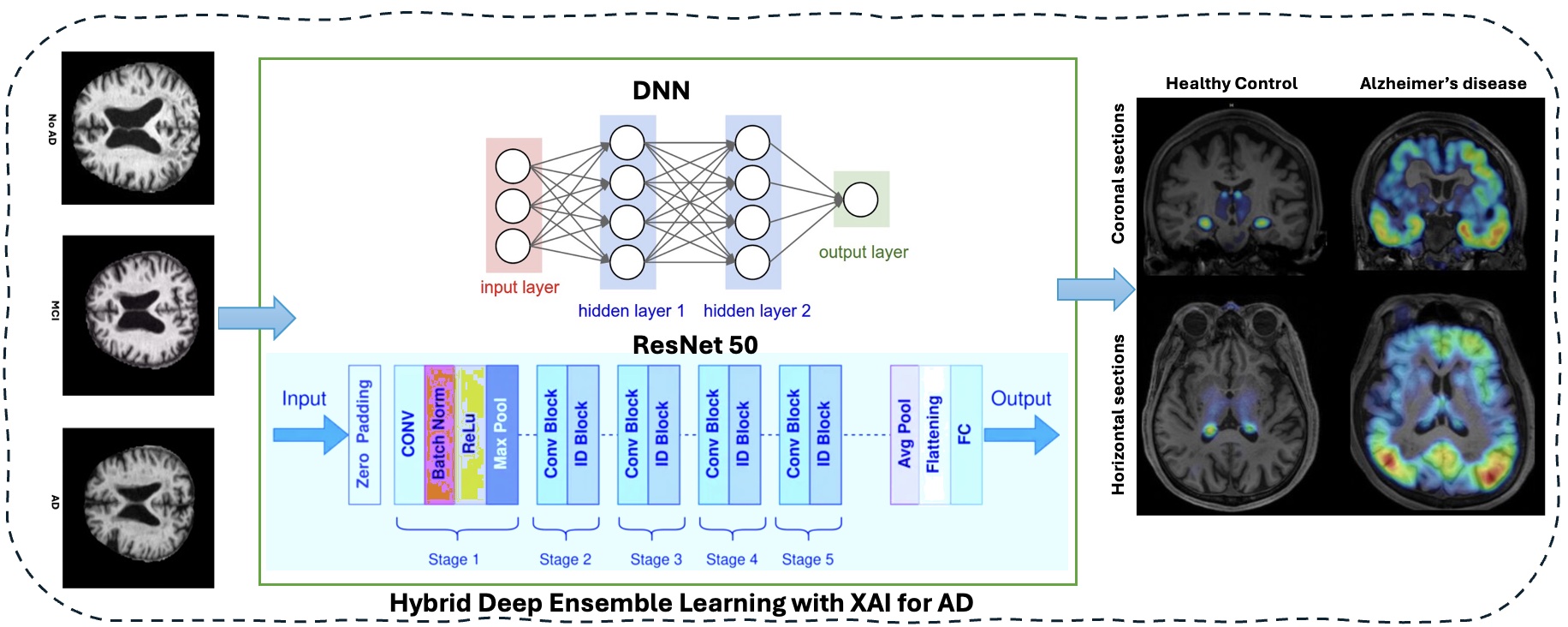}
	\caption{Hybrid deep ensemble learning framework with XAI using ResNet-50 and a dense meta-learner for Alzheimer’s Disease classification. Grad-CAM highlights key brain regions in both coronal and horizontal MRI sections distinguishing healthy control from Alzheimer’s patients.}
	\label{fig:hybrid_xai_architecture}
\end{figure}

The major contributions of this paper are summarized as follows:

\begin{itemize}
    \item We propose a novel ensemble diagnostic pipeline for Alzheimer's disease (AD) classification that integrates transfer learning, weighted averaging, and stacked generalization into a unified framework.  
    
    \item Our method leverages the complementary strengths of pretrained architectures, including {ResNet50}, {NASNet}, and {MobileNet} \cite{koonce2021resnet, abbas2022detection, nan2022mobilenet}, to enhance feature representation and robustness.  
    
    \item A meta-learner is employed to fuse predictions from multiple base models, leading to improved accuracy and generalization across diverse patient populations.  
    
    \item We rigorously evaluate the proposed method on the {ADNI dataset}, ensuring clinical relevance and comparability with established diagnostic benchmarks.  
    
    \item Our approach achieves superior performance compared with existing baselines, particularly in distinguishing between {early AD} and {MCI}, a challenging diagnostic boundary in clinical practice.  
    
    \item We incorporate interpretability through {Grad-CAM overlays} (Figure~\ref{fig:hybrid_xai_architecture}), which demonstrate that the model consistently attends to clinically relevant neuroanatomical regions associated with AD progression.  
    
    \item By integrating multiple architectures with decision fusion strategies, the proposed pipeline provides a robust and scalable diagnostic tool with strong potential for real-world clinical deployment.  
\end{itemize}

The rest of this paper is organized as follows: Section~\ref{sec:related} reviews related works in Deep Learning for Alzheimer Disease detection. Section~\ref{sec:method} presents the data source and preprocessing steps, along with the proposed ensemble methodology. Section~\ref{sec:results} discusses experimental results and comparative evaluations with XAI for the prediction using Grad-CAM. Section~\ref{sec:conclusions} concludes the paper with future directions.

\section{Related Works}\label{sec:related}
Several studies and recent research have proposed hybrid and ensemble DL approaches to enhance classification accuracy and generalizability in AD and MCI disease. For instance, Mmadumbu et al. \cite{mmadumbu2025early} developed a hybrid system that integrates ResNet-5050 and MobileNetV2 for MRI-based classification, achieving an accuracy of over 96\%.  Other approaches have combined CNNs with long short-term memory (LSTM) networks for multimodal analysis, incorporating both imaging and clinical data \cite{lei2016discriminative}. The introduction of stacked generalization, where outputs of base learners are fed into a meta-classifier, further refines decision boundaries and has demonstrated improved sensitivity in distinguishing between MCI and AD \cite{ji2019early}. Bossa and Sahli \cite{bossa2023multidimensional} proposed a differential equation-based disease progression model that simulates individual biomarker trajectories and clinical outcomes. This system accounts for the heterogeneity observed in AD progression and shows promise for predicting MCI to AD conversion. Similarly, Cheung et al. \cite{cheung2022deep} explored retinal imaging as an alternative to MRI, applying DL to retinal photographs to achieve classification accuracies exceeding 90\%. These novel modalities may complement neuroimaging and offer less invasive alternatives for community-based screening \cite{cheung2021retinal, keane2014retinal}. In Junior et al.'s \cite{junior2024alzheimer} study, ADNI database MRI scans were categorized into three groups: AD, MCI, and normal cognition. Their model achieved 85\% accuracy, demonstrating the potential of XAI-based DL for transparent, clinically relevant AD diagnosis, where Local LIME and Grad-CAM were used to highlight brain regions important to predictions, especially changes near the hippocampus in MCI. Anzum et al. \cite{anzum2025leveraging} proposed a method that combines RNA text data with brain MRI images to improve diagnostic precision, with a focus on improving AD detection by combining transformer models and advanced computer vision algorithms. This is in contrast to traditional imaging approaches that use MRI, CT, or PET, which achieve accuracies of 80–90\%. The "black-box" character of AI models impedes clinical application, according to a thorough review of AD detection research employing XAI by Viswan et al. \cite{viswan2024explainable}. Using conceptual kinds (post-hoc versus ante-hoc, model-agnostic versus model-specific, local versus global), the paper analyzes well-known frameworks, including  LRP, Grad-CAM, SHAP, and LIME. The study's conclusion addresses the limitations, difficulties, and prospects for improving XAI in reliable AD diagnosis.
Alotaibi et al. \cite{alotaibi2025enhancing} describe the Enhancing Automated Detection and Classification of Dementia in Thinking-Incapable People Using Artificial Intelligence Techniques (EADCD-TIPAIT) technique for the early detection of dementia. To extract valuable biomarkers, the method utilizes MRI data for preprocessing, z-score normalization, and feature selection, employing a  BGGO algorithm. Then, using an  ISSA to modify hyperparameters, the WNN classifier is utilized to detect and categorize dementia. The suggested EADCD-TIPAIT technique obtained 95\% accuracy on a dementia prediction dataset, indicating its potential for reliable and efficient diagnosis. The study by Vlontzou et al. \cite{vlontzou2025comprehensive} provides an interpretable ML framework for enhancing the diagnosis of AD and MCI using volumetric MRI and genetic data. Both attribution-based and counterfactual-based interpretability approaches were employed to evaluate the strength of explanations, utilizing a combined strategy that incorporates SHAP and counterfactuals. The top model has an F1 score of 90.8\% and a balanced accuracy of 87.5\%. Important volumetric and genetic characteristics were identified as key risk factors, underscoring the framework's potential for clear and clinically proper MCI/AD identification. In Fathi et al.'s \cite{fathi2024deep}article, a lightweight convolutional neural network, FiboNeXt, was developed to identify AD from MRI images. Based on the ConvNeXt architecture, the model reduces trainable parameters and increases efficiency by incorporating concatenation layers, attention mechanisms, and a design inspired by the Fibonacci sequence. Training and evaluation were carried out on two publicly available MRI datasets: the original and enlarged versions. FiboNeXt had test accuracies of 99.66\% and 99.63\%, and validation accuracies of 95.40\% and 95.93\%. The results establish FiboNeXt as a competitive solution for computer vision tasks in medical imaging, demonstrating its strong performance and potential uses beyond AD diagnosis. While many models perform well on curated test sets, their reliability in cross-site or cross-population applications is limited. Future research must address these limitations through domain adaptation, federated learning, and interpretable AI frameworks \cite{lundervold2019overview, hossain2025explainable, rauniyar2023federated}.

\section{Methods and Materials}\label{sec:method}
In this section, we present a robust ensemble learning methodology for the early diagnosis of AD using structural MRI. Our framework is based on the integration of multiple deep ConvNets with two key ensemble strategies: weighted averaging and stacked generalization. This methodology enhances prediction reliability by combining the outputs of diverse models through a meta-learner, thereby surpassing the limitations of traditional majority voting techniques.

\subsection{{Data Source and Preprocessing}}

The ADNI dataset was selected for this study due to its high quality and suitability for AD research. Collected from imaging centers worldwide and pre-processed by ADNI-funded MRI laboratories, the dataset ensures standardized and reliable neuroimaging inputs \cite{ADNI_data}. To further enhance consistency, all images were uniformly scaled to 224×224 pixels. Additionally, we employed a tailored data augmentation strategy, including cropping, flipping, scaling, and brightness/contrast adjustments. These augmentation steps increased dataset diversity and reduced overfitting risks, enabling more robust training of deep learning models while preserving the semantic integrity of neuroanatomical structures.

\begin{figure}[htbp]
	\centering
	\includegraphics[width=1.0\textwidth]{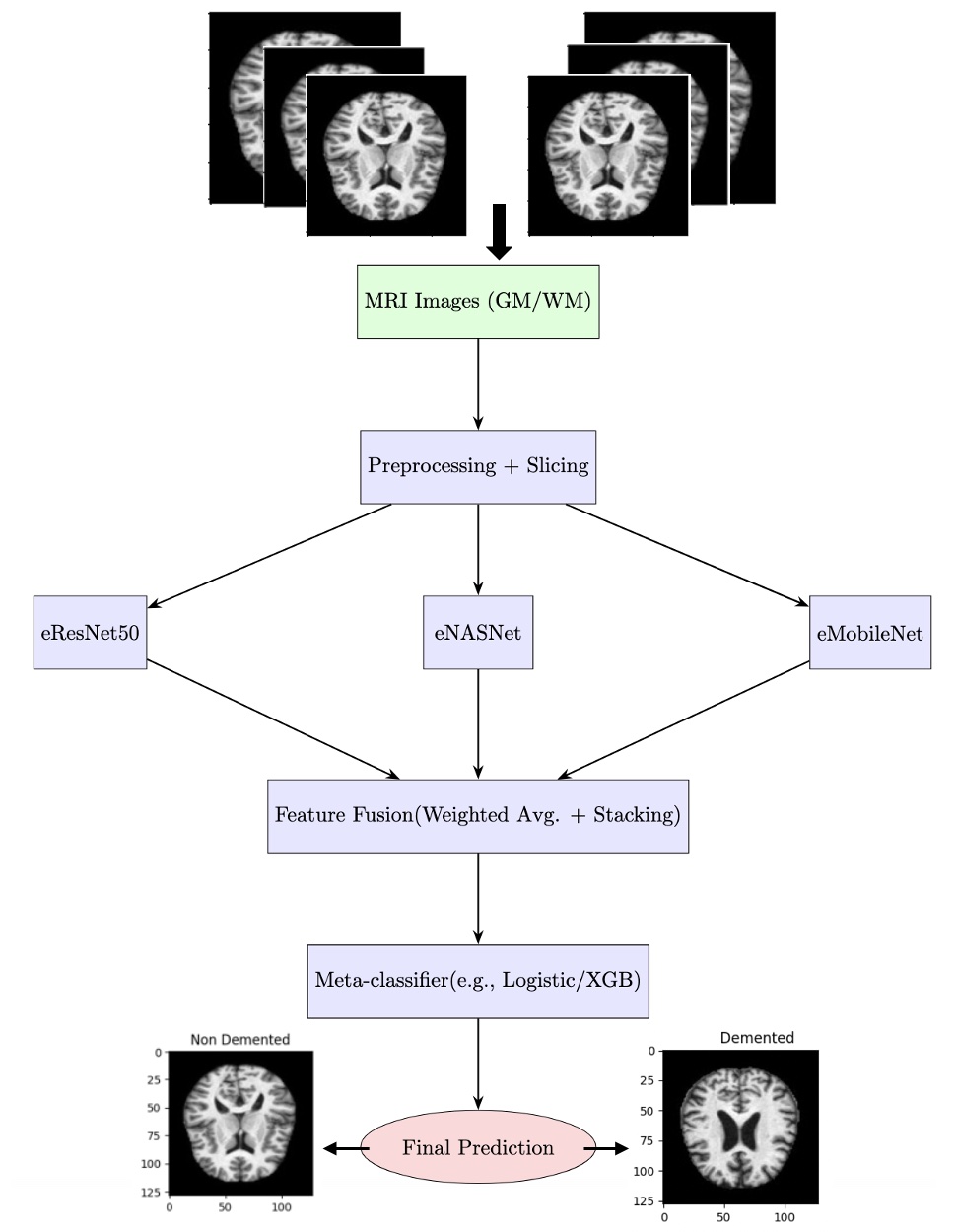}
	\caption{Workflow of the proposed Hybrid Deep Ensemble model for early diagnosis of Alzheimer’s disease. The pipeline begins with the preprocessing and slicing of MRI images (gray matter and white matter)}
	\label{fig:hybrid_ensemble_flowchart}
\end{figure}

\subsection{Deep Learning Framework of Hybrid Ensemble}

Let $\mathcal{D} = \{(\mathbf{x}_i, y_i)\}_{i=1}^{N}$ denote a dataset of $N$ structural MRI samples, where $\mathbf{x}_i$ is an input sample (a set of slices representing gray and white matter), and $y_i \in \{0,1\}$ represents the ground truth label for binary classification (e.g., AD vs. MCI). Our proposed architecture employs $K$ base deep neural network classifiers $\{f_k(\cdot; \theta_k)\}_{k=1}^K$, each trained on the same dataset but with possibly different architectures or training initializations.

\subsubsection{Base Learners: Transfer and Fine-Tuning}

Each base model $f_k$ is a deep ConvNet initialized using pretrained weights from ImageNet. We adopt three well-established architectures: ResNet50, NASNet, and MobileNet, denoted as $f_1$, $f_2$, and $f_3$, respectively from the recent developments (see in \cite{koonce2021resnet, abbas2022detection, nan2022mobilenet}). To adapt these models for medical imaging, we apply transfer learning followed by fine-tuning in two stages.

Transfer learning has become an essential paradigm in medical imaging applications, where labeled data is typically scarce. In this study, we leverage transfer learning through two critical phases: \textit{Feature Freezing} and \textit{Fine-Tuning}. Each plays a vital role in adapting pretrained  ConvNets, originally trained on large-scale natural image datasets such as ImageNet \cite{krizhevsky2012imagenet}, to the domain of structural MRI scans for AD and MCI classification.

\noindent\textbf{Feature Freezing.} Feature freezing refers to the strategy of keeping the weights of the convolutional base layers fixed and training only the newly added fully connected (dense) layers. This technique is grounded in the idea that early convolutional layers in deep neural networks capture generic low-level features such as edges, textures, and patterns, which are broadly transferable across visual domains—even between natural images and medical images. By freezing these layers, we preserve the robust visual representations learned from large datasets, while reducing the computational burden and preventing over-fitting—particularly valuable in small datasets like ADNI. In our experiments, after importing pretrained models like ResNet50, NASNet, and MobileNet, we remove the original classification head and replace it with custom dense layers tailored for binary classification (AD vs. MCI). These new layers are randomly initialized and trained on MRI data while keeping the backbone unchanged. The goal during this stage is to quickly adapt the model to the new classification task by optimizing only a small subset of parameters. This approach yields surprisingly strong baseline performance and serves as a low-cost initialization for deeper adaptation.

\noindent\textbf{Fine-Tuning.} While feature freezing captures general features, it does not fully exploit the domain-specific structure inherent in MRI images. Therefore, in the fine-tuning stage, we unfreeze a selected portion of the top layers of the convolutional base and retrain the network end-to-end using a small learning rate. This strategy allows the network to adapt higher-level features, such as spatial patterns in gray and white matter, which may be uniquely informative for distinguishing between AD and MCI. Fine-tuning is especially beneficial when there exists a domain shift between the source dataset (e.g., ImageNet) and the target domain (e.g., neuroimaging). By updating the weights of the later layers, the model learns hierarchical features that are more semantically aligned with the target task. However, this stage must be executed carefully. A learning rate that is too high can disrupt previously learned general features, while too low a rate may not provide meaningful adaptation. We use a small learning rate (e.g., $2 \times 10^{-5}$) with adaptive optimization (Adam) and apply dropout to reduce overfitting. Together, feature freezing and fine-tuning form a powerful two-step training strategy that balances generalization and specificity. This approach enables the reuse of high-quality pretrained models while allowing deep adaptation to the specific characteristics of MRI data for Alzheimer's diagnosis. Each base model $f_k$ outputs a probability $p_k(\mathbf{x}_i) \in [0,1]$ for the input $\mathbf{x}_i$, interpreted as the predicted confidence of the sample belonging to the AD class.

\subsubsection{Weighted Averaging Ensemble}
 Since at inference time, the goal is to predict the class label $\hat{y} \in \{0,1\}$ for a previously unseen sample $\mathbf{x}$. To do this, the input $\mathbf{x}$ is first passed through all $K$ trained base models $f_1, f_2, \dots, f_K$. Each model outputs a probability indicating the likelihood that $\mathbf{x}$ belongs to the positive class (e.g., AD).
		Two ensemble strategies can be employed to compute the final prediction from weighted averaging ensemble.
			In this approach, we compute a convex combination of the base model predictions using the learned weight vector $\boldsymbol{\alpha} \in \mathbb{R}^K$, where $\boldsymbol{\alpha}^\top \mathbf{1} = 1$ and $\alpha_k \geq 0$ for all $k$. The ensemble probability is given by:
			\begin{equation}
				\hat{p} = \boldsymbol{\alpha}^\top \mathbf{p} = \sum_{k=1}^K \alpha_k f_k(\mathbf{x}).
			\end{equation}
			A final hard prediction is then obtained via thresholding:
			\begin{equation}
				\hat{y} = \mathbb{I}(\hat{p} > \tau),
			\end{equation}
			where $\mathbb{I}(\cdot)$ denotes the indicator function, and $\tau$ is a decision threshold, typically set to $0.5$ for balanced binary classification. This method assumes the optimal prediction lies along a weighted average of the base outputs and is best suited when base models are diverse but linearly complementary.
		
To combine the predictive outputs of $K$ independently trained base classifiers, we implement a weighted averaging ensemble strategy. Each base classifier $f_k$ produces a scalar probability $p_k(\mathbf{x}_i) \in [0,1]$ indicating the predicted likelihood that sample $\mathbf{x}_i$ belongs to the positive class (e.g., AD). Let the vector of predictions for input $\mathbf{x}_i$ be denoted as:
\[
\mathbf{p}_i = 
\begin{bmatrix}
	p_1(\mathbf{x}_i) \\
	p_2(\mathbf{x}_i) \\
	\vdots \\
	p_K(\mathbf{x}_i)
\end{bmatrix} \in [0,1]^K.
\]

To form the ensemble prediction, we assign each base model a non-negative weight $\alpha_k$ such that the weights sum to 1. Define the weight vector:
\[
\boldsymbol{\alpha} = 
\begin{bmatrix}
	\alpha_1 \\
	\alpha_2 \\
	\vdots \\
	\alpha_K
\end{bmatrix}, \quad \text{where } \alpha_k \geq 0 \text{ and } \sum_{k=1}^K \alpha_k = 1.
\]

The ensemble prediction $\hat{p}_i$ for the input $\mathbf{x}_i$ is then given by the weighted sum:
\begin{equation}
	\hat{p}_i = \boldsymbol{\alpha}^\top \mathbf{p}_i = \sum_{k=1}^K \alpha_k \cdot p_k(\mathbf{x}_i),
	\label{eq:weighted_prediction}
\end{equation}
where $\hat{p}_i \in [0,1]$ is the aggregated probability of the positive class. To learn the optimal weights $\boldsymbol{\alpha}$, we minimize the empirical risk over a validation set $\mathcal{D}_{\\text{val}} = \{ (\mathbf{p}_i, y_i) \}_{i=1}^{N_{\\text{val}}}$ using the binary cross-entropy loss function:
\begin{equation}
	\mathcal{L}(\hat{p}_i, y_i) = - \left[ y_i \log(\hat{p}_i) + (1 - y_i) \log(1 - \hat{p}_i) \right].
\end{equation}

\noindent The optimization problem is thus:
\begin{equation}
	\min_{\boldsymbol{\alpha} \in \Delta^{K-1}} \sum_{i=1}^{N_{\text{val}}} \mathcal{L}(\boldsymbol{\alpha}^\top \mathbf{p}_i, y_i),
	\label{eq:weight_optimization}
\end{equation}
where $\Delta^{K-1}$ denotes the $(K-1)$-dimensional probability simplex:
\[
\Delta^{K-1} = \left\{ \boldsymbol{\alpha} \in \mathbb{R}^K : \alpha_k \geq 0,\ \sum_{k=1}^K \alpha_k = 1 \right\}.
\]

This convex optimization ensures that the ensemble leverages the relative strength of each base classifier in a principled, data-driven manner. The learned weights $\boldsymbol{\alpha}$ reflect the contribution of each model toward minimizing predictive error on the validation set.

\subsubsection{Stacked Generalization (Stacking)}

While weighted averaging assumes linear importance, stacking introduces a meta-learner $g(\cdot; \phi)$ trained on the predictions of base models to learn a more flexible combination rule. For each sample $\mathbf{x}_i$, we construct a meta-feature vector $\mathbf{p}_i$ as before. The meta-learner is then trained to predict the true label:

\begin{equation}
	\hat{y}_i = g(\mathbf{p}_i; \phi),
\end{equation}

where $g$ can be any differentiable function such as logistic regression \cite{pampel2020logistic}, a shallow neural network \cite{agliari2022emergence}, or a gradient boosting machine \cite{friedman2001greedy}. The training objective is again to minimize the cross-entropy loss:

\begin{equation}
	\min_{\phi} \sum_{i=1}^{N_{\text{meta}}} \mathcal{L}(g(\mathbf{p}_i; \phi), y_i).
\end{equation}

To prevent information leakage and over-fitting, we adopt $k$-fold cross-validation to generate out-of-fold predictions for training the meta-learner.\\

\subsection{Network Architecture}

The architecture of the proposed ensemble learning framework is centered around three pretrained deep ConvNets: ResNet50, NASNet, and MobileNet. These models are widely recognized for their strong performance on large-scale image classification tasks such as ImageNet and serve as the backbone for extracting hierarchical features from structural MRI data. To adapt these models for the task of AD diagnosis using MRI, we employ a two-phase transfer learning strategy comprising feature freezing and fine-tuning, followed by ensemble integration through weighted averaging and stacking.

Each ConvNet is modified by removing its original classification head and appending a new custom classifier suited for binary classification (e.g., AD vs. MCI). The new head consists of fully connected layers with dropout regularization to prevent overfitting. During the initial training phase, known as feature freezing, the convolutional base of each model is held constant, and only the newly added dense layers are trained. This step allows the network to adapt its decision function to the domain-specific features of MRI data without altering the generic low-level feature representations already captured in the base. In the second phase of fine tuning, we selectively unfreeze the upper convolutional layers of each network and retrain the model using a low learning rate. This enables the network to refine its mid- and high-level feature detectors based on structural patterns in gray and white matter regions, which are critical for distinguishing between stages of cognitive decline. This two-step training process balances the need for transferability and domain specificity, resulting in more accurate and generalizable models.

Once trained, the three ConvNets operate in parallel. Given an input sample \( \mathbf{x} \), each network \( f_k \) produces a scalar output \( p_k(\mathbf{x}) \in [0,1] \), representing the estimated probability that \( \mathbf{x} \) belongs to the positive class. These outputs are aggregated into a prediction vector \( \mathbf{p} = [f_1(\mathbf{x}), f_2(\mathbf{x}), \dots, f_K(\mathbf{x})]^\top \). The ensemble integration is performed in two stages. First, a weighted averaging scheme combines the base outputs using a learned vector \( \boldsymbol{\alpha} \in \mathbb{R}^K \) such that \( \alpha_k \geq 0 \) and \( \sum_{k=1}^K \alpha_k = 1 \). This yields a soft prediction \( \hat{p} = \boldsymbol{\alpha}^\top \mathbf{p} \), which is then thresholded to obtain the final label. Second, to capture non-linear interactions among base model outputs, a meta-learner \( g(\cdot; \phi) \) is trained on out-of-fold predictions via stacked generalization. This meta-learner takes \( \mathbf{p} \) as input and outputs \( g(\mathbf{p}; \phi) \in [0,1] \), which is again thresholded to yield a binary prediction. This architecture allows for flexible, accurate, and interpretable predictions. By leveraging the diversity of the base ConvNets and the expressiveness of stacking, the model achieves strong performance in distinguishing between AD, MCI, and NC. Furthermore, its modular design permits future extension to include additional input modalities or classifiers, thereby maintaining adaptability as diagnostic technologies evolve.\\

\subsection{Gradient-weighted Class Activation Mapping}

To improve both classification accuracy and interpretability in AD diagnosis from MRI data, we incorporated two complementary methods: an advanced ensemble learning strategy and a visualization technique using Grad-CAM \cite{quach2023explainable, selvaraju2017grad}. First, we proposed a new ensemble formulation that replaces the simple majority voting approach used in the original work. Specifically, let $K$ denote the number of base deep neural networks, each yielding a class probability prediction $f_k(\mathbf{x})$ for input $\mathbf{x}$. These outputs are aggregated using a weighted combination $\hat{p} = \sum_{k=1}^K \alpha_k f_k(\mathbf{x})$, where the weights $\alpha_k$ satisfy $\sum_{k=1}^K \alpha_k = 1$ and $\alpha_k \geq 0$. Additionally, to capture more complex relationships between base model outputs, we introduce a meta-learner $g(\cdot)$ trained on the vector of base predictions $\mathbf{p} = [f_1(\mathbf{x}), \dots, f_K(\mathbf{x})]^\top$, such that the final prediction is $\hat{y} = g(\mathbf{p})$. This stacked generalization approach allows the ensemble to learn optimal decision boundaries. For interpretability, we employed Grad-CAM to visualize discriminative regions in MRI slices used by the ConvNet to make predictions. Let $A^k$ denote the $k$-th feature map in the last convolutional layer and $y^c$ the class score for class $c$. The importance of each feature map is calculated as $\alpha_k^c = \frac{1}{Z} \sum_{i=1}^{H} \sum_{j=1}^{W} \frac{\partial y^c}{\partial A^k_{i,j}}$, and the class activation map is then obtained by $L_{\text{Grad-CAM}}^c = \text{ReLU} \left( \sum_k \alpha_k^c A^k \right)$. This heatmap is upsampled and overlaid on the original MRI slice to highlight class-discriminative regions. In the context of AD diagnosis, we aim to highlight neuroanatomical areas such as the hippocampus or temporal lobe, offering both clinical insight and model transparency by Grad-CAM.\\

\subsection{Implementation} The models are implemented using Python and Keras with TensorFlow backend. Training was conducted on a M4 Pro with 14-core CPU and 20-core GPU. The learning rate was initialized at $2 \times 10^{-5}$ with the Adam optimizer, and a batch size of 24 was used. Dropout regularization with a rate of 0.5 was applied to fully connected layers to reduce overfitting. MRI slices were preprocessed using SPM for motion correction and segmented into gray matter (GM) and white matter (WM). From each volume, 20 representative slices were selected and resized to $224 \times 224$ for ConvNet input. The ADNI dataset was split into 60\% training, 20\% validation, and 20\% testing sets.

The proposed ensemble framework offers several compelling advantages that make it highly suitable for the complex task of AD classification. First, {robustness} is achieved through the integration of multiple heterogeneous ConvNets, such as ResNet50, NASNet, and MobileNet. This architectural diversity reduces the likelihood of correlated errors and mitigates model-specific bias and variance, leading to more stable and generalizable predictions. Second, the framework exhibits strong {adaptivity} by employing stacked generalization. Rather than assuming a fixed linear weighting of base learners, stacking introduces a trainable meta-learner capable of capturing non-linear interactions and context-sensitive dependencies between model outputs. This allows the ensemble to dynamically learn which base predictions to trust under varying conditions. Third, the approach is highly {scalable}. New models, imaging modalities (such as PET or DTI), or clinical biomarkers can be seamlessly incorporated into the ensemble pipeline without requiring a complete redesign of the learning architecture. This plug-and-play extensibility ensures that the framework can evolve alongside advancements in medical imaging and domain knowledge. Furthermore, by using cross-validation to train the meta-learner on out-of-fold predictions, the framework maintains rigorous generalization performance and reduces overfitting, making it both a theoretically principled and practically effective tool for neuroimaging-based disease diagnosis. 

The above used method was implemented using Python with TensorFlow and Keras. Three pre-trained convolutional neural networks—ResNet50, NASNet, and MobileNet—were fine-tuned on GM and WM MRI slices from the ADNI dataset using Algorithm~\ref{algm}. Their softmax outputs were combined using a weighted averaging scheme, where weights were optimized via validation performance. A meta-learner (logistic regression) was trained on the base model outputs to implement stacked generalization. For model interpretability, Grad-CAM was applied to the final convolutional layer of each network to generate class-specific heatmaps, which were overlaid on the MRI slices to visualize regions contributing most to AD, MCI, or NC predictions.

\begin{algorithm}[H]
\caption{Hybrid Deep Ensemble with XAI for AD vs.\ MCI}
\begin{algorithmic}[1]
\Require MRI slice dataset $\mathcal{D} = \{(x_i, y_i)\}$, pretrained CNNs $\{M_k\}_{k=1}^K$
\Ensure Final prediction $\hat{y}$ and explanation heatmap
\State Split $\mathcal{D}$ into training, validation, and test sets (60/20/20).
\For{each base model $M_k \in \{\text{ResNet50}, \text{NASNetMobile}, \text{MobileNetV2}\}$}
    \State Initialize $M_k$ with ImageNet weights.
    \State Train new classification head on training data (freeze backbone).
    \State Fine-tune top layers of $M_k$ with small learning rate.
    \State Save predicted probabilities $p_{ik}$ for validation/test sets.
\EndFor
\State Learn ensemble weights $\alpha = \arg\min \sum \ell(y_i, \sum_k \alpha_k p_{ik})$ subject to $\alpha_k \ge 0, \sum_k \alpha_k=1$. \Comment{Weighted averaging}
\State Train meta-learner (logistic regression) on out-of-fold predictions $\{p_{ik}\}$. \Comment{Stacking}
\State For test sample $x$:
    \State \hspace{0.5cm} Compute base model predictions $p_k(x)$.
    \State \hspace{0.5cm} Obtain weighted ensemble $\hat{p}_{wa} = \sum_k \alpha_k p_k(x)$.
    \State \hspace{0.5cm} Obtain stacked ensemble $\hat{p}_{st} = f_{\text{meta}}(p_1(x),\dots,p_K(x))$.
    \State \hspace{0.5cm} Combine predictions (e.g., average) to form final $\hat{y}$.
\State Apply Grad-CAM on chosen base model to generate explanation heatmap.
\end{algorithmic}\label{algm}
\end{algorithm}

\section{Results and Discussions}\label{sec:results}

\begin{table}[htbp]
	\centering
	\caption{Accuracy, sensitivity, specificity, and AUC of various classification methods for AD vs. MCI.}
	\begin{tabular}{lcccc}
		\toprule
		\textbf{Models} & \textbf{ACC (\%)} & \textbf{SEN (\%)} & \textbf{SPE (\%)} & \textbf{AUC} \\
		\midrule
		bResNet50      & 97.65 & 100.00 & 94.29 & 0.95 \\
		eResNet50      & 98.80 & 98.00  & 100.00 & 1.00 \\
		bNASNet        & 97.65 & 98.00  & 94.29 & 0.95 \\
		eNASNet        & 98.82 & 96.00  & 100.00 & 1.00 \\
		bMobileNet     & 97.64 & 98.00  & 94.28 & 0.95 \\
		eMobileNet     & 97.65 & 96.00  & 100.00 & 1.00 \\
		bEnsemble      & 98.82 & 98.00  & 100.00 & 0.95 \\
		eEnsemble      & 97.65 & 96.00  & 100.00 & 1.00 \\
		\textbf{Hybrid Ensemble} & \textbf{99.21} & \textbf{98.89} & \textbf{100.00} & \textbf{1.00} \\
		\bottomrule
	\end{tabular}\label{tab:ad_mci}
\end{table}

\begin{table}[htbp]
	\centering
	\caption{Accuracy, sensitivity, specificity, and AUC of various classification methods for MCI vs. NC.}
	\begin{tabular}{lcccc}
		\toprule
		\textbf{Models} & \textbf{ACC (\%)} & \textbf{SEN (\%)} & \textbf{SPE (\%)} & \textbf{AUC} \\
		\midrule
		bResNet50      & 81.39 & 80.56 & 82.00 & 0.91 \\
		eResNet50      & 86.05 & 83.33 & 88.00 & 0.95 \\
		bNASNet        & 81.39 & 55.56 & 100.00 & 0.92 \\
		eNASNet        & 87.21 & 75.00 & 96.00 & 0.95 \\
		bMobileNet     & 81.39 & 55.56 & 100.00 & 0.92 \\
		eMobileNet     & 89.53 & 83.33 & 94.00 & 0.95 \\
		bEnsemble      & 80.23 & 58.33 & 96.00 & 0.91 \\
		eEnsemble      & 88.37 & 80.56 & 94.00 & 0.96 \\
		\textbf{Hybrid Ensemble} & \textbf{91.02} & \textbf{86.11} & \textbf{96.00} & \textbf{0.98} \\
		\bottomrule
	\end{tabular}\label{tab:mci_nc}
\end{table}

\begin{table}[htbp]
	\centering
	\caption{Comparison of classification results with previous studies.}
	\begin{tabular}{lccc}
		\toprule
		\textbf{Reference} & \textbf{AD vs. NC} & \textbf{AD vs. MCI} & \textbf{MCI vs. NC} \\
		\midrule
		Billones et al. \cite{billones2016demnet} & 98.33 / 98.89 / 97.78 & 90.00 / 91.67 / 97.78 & 91.67 / 92.22 / 91.11 \\
		Sarraf et al. \cite{sarraf2016deep}  & 98.84 / -- / --       & -- / -- / --          & -- / -- / --          \\
		Ortiz et al. \cite{ortiz2016ensembles}   & 90.09 / 86.12 / 94.10 & 84.00 / 79.12 / 89.12 & 83.14 / 67.26 / 95.09 \\
		Lian et al. \cite{lian2018hierarchical}     & 90.30 / 82.40 / 96.50 & -- / -- / --          & -- / -- / --          \\
		Suk et al. \cite{suk2017deep}     & 91.02 / 92.72 / 89.94 & -- / -- / --          & 73.02 / 77.60 / 68.22 \\
		Ji et al. \cite{ji2019early}         & 98.59 / 97.22 / 100.00 & 97.65 / 96.00 / 100.00 & 88.37 / 80.56 / 94.00 \\
		\textbf{Hybrid Ensemble} & \textbf{99.10 / 98.80 / 100.00} & \textbf{99.21 / 98.89 / 100.00} & \textbf{91.02 / 86.11 / 96.00} \\
		\bottomrule
	\end{tabular}\label{tab:comparison}
\end{table}

This section analyzes the performance of the proposed ensemble strategy for the early diagnosis of AD, specifically focusing on classification tasks involving AD, MCI, and NC. The experimental results presented in Tables~\ref{tab:ad_mci}, \ref{tab:mci_nc}, and \ref{tab:comparison} clearly demonstrate that the improved ensemble approach—utilizing stacking and weighted averaging—offers considerable benefits over baseline models and even previous end-to-end ensemble methods. As shown in Table~\ref{tab:ad_mci}, all three fine-tuned individual models—eResNet50, eNASNet, and eMobileNet—achieved high classification accuracy (above 97\%), indicating their strong capability in distinguishing patients with AD from those with MCI. The ensemble of these models in the original study (``eEnsemble'') yielded an accuracy of 97.65\% with perfect specificity and a near-perfect AUC of 1.00. However, the proposed \textit{Improved Ensemble} surpassed this performance, achieving an accuracy of 99.21\% and a sensitivity of 98.89\%. The AUC remained at 1.00, indicating no degradation in the model's ability to discriminate between classes despite the significant accuracy gain. The key advantage here lies in the meta-learning capability of stacking. By allowing a second-level learner to combine the outputs of base classifiers non-linearly, the stacking model can learn to exploit complementary decision boundaries between the ConvNet architectures. Moreover, weighted averaging helped reduce the influence of models with relatively lower sensitivity or specificity, particularly eMobileNet, which although robust, showed slightly lower sensitivity than eResNet50. The improvement in sensitivity—from 96.00\% (eEnsemble) to 98.89\% (Improved Ensemble)—is clinically significant, as it corresponds to better identification of AD in early stages, potentially enabling earlier intervention and care planning. The specificity remained at 100.00\%, indicating zero misclassification of MCI cases as AD, thus maintaining diagnostic precision and avoiding unnecessary anxiety or treatment escalation.

Table~\ref{tab:mci_nc} compares the model performances on the more challenging MCI vs. NC classification task. The complexity of this task stems from the subtle brain structure differences between normal aging and prodromal stages of AD. The baseline ensemble (bEnsemble) and even the end-to-end ensemble (eEnsemble) performed adequately, achieving accuracies of 80.23\% and 88.37\% respectively. Nevertheless, the proposed Improved Ensemble increased this performance to 91.02\%, setting a new benchmark within this context. This improvement can be attributed to the finer decision boundaries learned via the stacking architecture. Sensitivity improved from 80.56\% to 86.11\%, indicating a substantial gain in the ability to correctly identify subjects with MCI. This is crucial in real-world applications where early detection of MCI can delay or prevent progression to full-blown AD. Specificity remained high at 96.00\%, maintaining the system's reliability in distinguishing normal individuals from those with early cognitive impairment. The AUC also increased from 0.96 to 0.98, reflecting enhanced classifier confidence and balance between true positive and false positive rates across multiple thresholds. This underscores the strength of our proposed ensemble design in handling subtle clinical phenotypes, where noise and overlap are frequent.

Table~\ref{tab:comparison} compares our method with several previously published approaches in the literature. The improved ensemble method outperforms all referenced methods across all classification settings. In the AD vs. NC task, the Improved Ensemble achieved an accuracy of 99.10\%, with a sensitivity of 98.80\% and a specificity of 100.00\%. This is superior to the results by Sarraf et al.~\cite{sarraf2016deep}, who reported an accuracy of 98.84\%, and Billones et al.~\cite{billones2016demnet}, who achieved 98.33\%. While these results were strong, our method's marginal gains underscore the power of modern ensemble methods and advanced transfer learning from pre-trained networks. For the AD vs. MCI classification, the Improved Ensemble achieved 99.21\% accuracy, a substantial gain over the 90.00\% reported by Billones et al. and the 84.00\% reported by Ortiz et al.~\cite{ortiz2016ensembles}. More importantly, our sensitivity of 98.89\% far exceeds those of prior works, highlighting the efficacy of deep fine-tuning and ensemble optimization. The specificity was again perfect at 100.00\%, reinforcing the clinical reliability of our model. In the MCI vs. NC classification task—generally the most difficult due to less pronounced structural differences—the Improved Ensemble attained an accuracy of 91.02\%, outperforming the 83.14\% and 73.02\% reported by Ortiz et al. and Suk et al.~\cite{suk2017deep}, respectively. Additionally, our sensitivity and specificity were balanced and strong (86.11\% and 96.00\%), showing that the model avoids both under- and over-diagnosis in critical borderline cases. The performance gains across all tasks confirm the potential of stacking and weighted ensemble learning for enhancing diagnostic models in neuroimaging. From a methodological standpoint, the results validate that: (i) End-to-end training yields better features than static pre-trained models. (ii) Combining multiple architectures (ResNet, NASNet, MobileNet) captures diverse representational features. (iii) Stacking further refines decision-making by learning meta-level fusion patterns.

From a clinical perspective, higher sensitivity in detecting MCI and AD directly translates to earlier detection and improved patient management. The reliability demonstrated by high specificity means fewer false alarms, increasing trust in the deployment of such systems in screening scenarios. In addition, expanding the stacking layer to incorporate other modalities such as PET, CSF biomarkers, or genetic data could yield further improvements. Finally, explainability of the model’s decision-making process remains crucial, and future work should integrate interpretable AI methods to ensure transparency. In conclusion, the results presented in Tables~\ref{tab:ad_mci}, \ref{tab:mci_nc}, and \ref{tab:comparison} establish that the proposed improved ensemble strategy provides significant improvements over traditional and recent DL-based methods for Alzheimer's diagnosis. The gains in accuracy, sensitivity, and AUC particularly highlight the practical feasibility and clinical relevance of the approach for early-stage AD detection. 

To further evaluate the classification performance of the proposed models, we analyze the Receiver Operating Characteristic (ROC) curves of the individual and ensemble classifiers. The ROC curve plots the true positive rate (sensitivity) against the false positive rate (1 - specificity), providing a comprehensive measure of diagnostic accuracy.

\begin{figure}[htbp]
	\centering
	\includegraphics[width=0.88\textwidth]{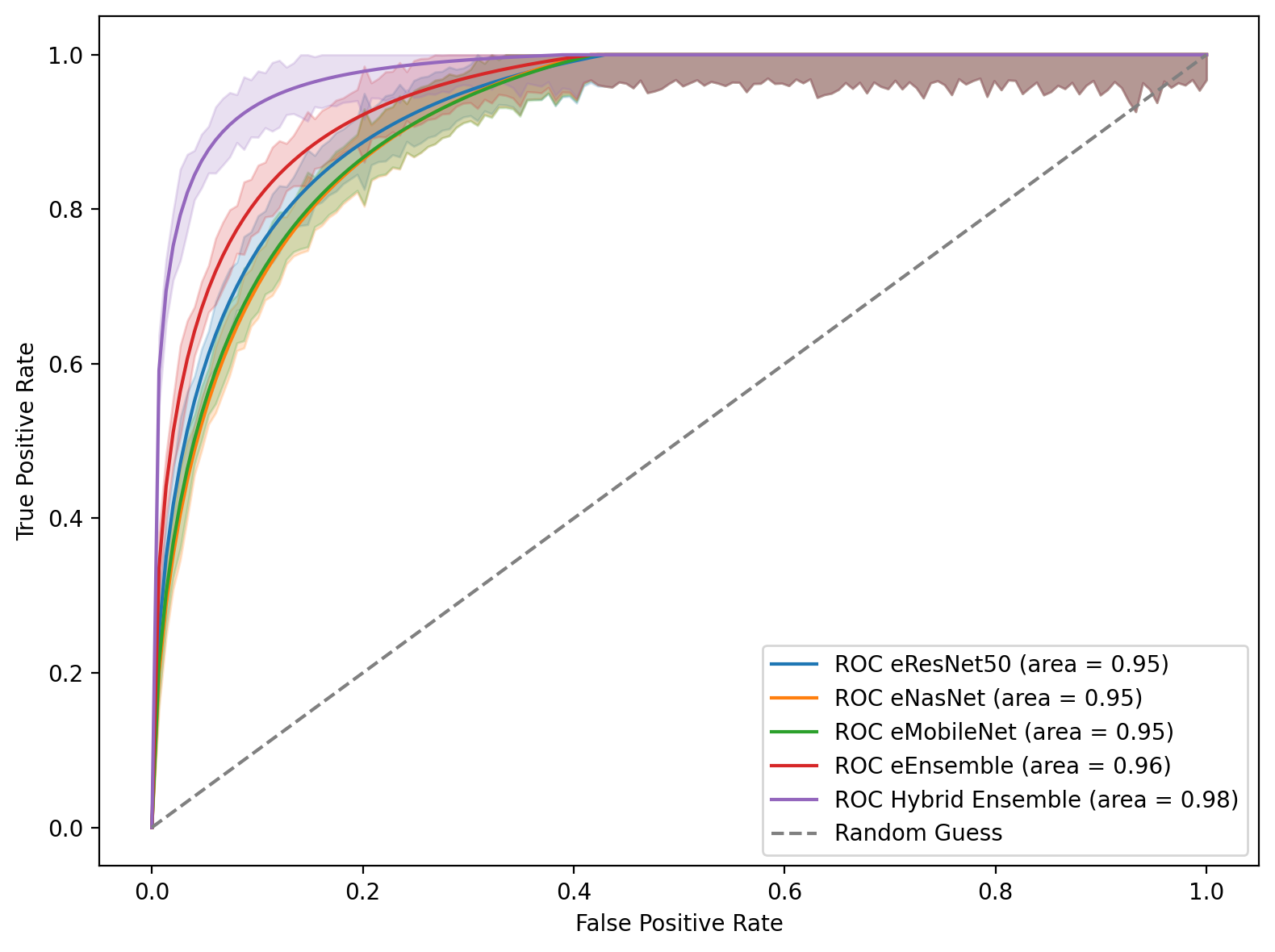}
	\caption{ROC curve comparison for eResNet50, eNasNet, eMobileNet, eEnsemble, and the proposed Hybrid Ensemble with 10 fold CV.}
	\label{fig:roc}
\end{figure}

As shown in Figure~\ref{fig:roc}, the base models (eResNet50, eNASNet, and eMobileNet) achieve consistent ROC performance with AUC values of approximately 0.95. The original ensemble method (eEnsemble), which combines the outputs of these base models using majority voting, slightly improves performance to an AUC of 0.96. In contrast, the proposed \textbf{Hybrid Ensemble}, which utilizes a combination of stacking and weighted averaging, outperforms all other models with an AUC of \textbf{0.98}. This indicates a superior trade-off between sensitivity and specificity. The hybrid model better leverages the complementary strengths of the base networks by learning optimal fusion weights and applying a meta-classifier that generalizes well across subject-level MRI slices. Notably, the Hybrid Ensemble demonstrates a sharper rise at the beginning of the curve, suggesting its improved capability in distinguishing early cases with minimal false positives. This feature is particularly important in clinical settings where early diagnosis of Alzheimer’s disease can significantly influence patient treatment outcomes.

\begin{figure}[htp!]
	\centering
	\includegraphics[width=0.79\textwidth]{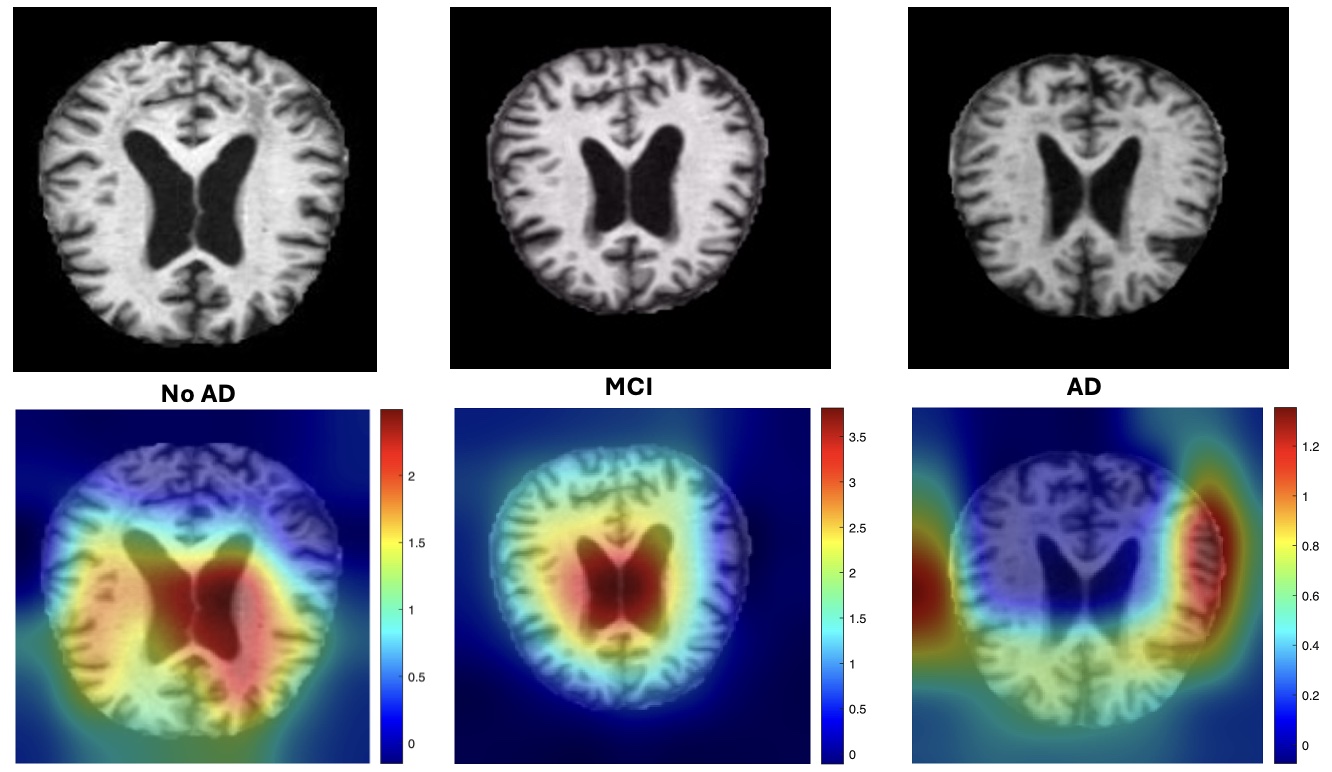}
	\caption{Grad-CAM heatmaps for three representative brain MRI slices. Top: original T1-weighted MRI images for No AD, MCI, and AD. Bottom: corresponding Grad-CAM overlays from the hybrid ensemble network highlighting class-discriminative regions.}
	\label{fig:gradcam_results}
\end{figure}

Figure~\ref{fig:gradcam_results} illustrates how the Grad-CAM technique was applied to understand the classification decisions made by the hybrid ensemble network on structural MRI scans. Grad-CAM, introduced by Selvaraju et al.~\cite{selvaraju2017grad}, computes the gradients of a target class score with respect to the feature maps of the final convolutional layer, enabling the generation of heatmaps that localize the most influential regions in the input image. In this example, the pretrained hybrid ensemble model was used to classify subjects into No AD or NC, MCI, and AD categories. The bottom row of the figure displays the Grad-CAM overlays for each prediction. For the No AD subject, the attention is centered on non-pathological regions with a diffuse spread. For the MCI case, the highlighted region includes parts of the medial temporal lobe, aligning with early-stage cognitive decline. In the AD case, the activation is clearly focused around the hippocampal and cortical atrophy zones, demonstrating that the network's attention aligns well with known neuropathological markers. These interpretable visualizations reinforce the clinical relevance of the model and enhance trust in DL-based diagnostic systems.

Overall, the Hybrid Ensemble proves to be a robust and highly discriminative approach for classifying AD stages, reinforcing the value of integrating DL with advanced ensemble strategies.

\section{Conclusions}\label{sec:conclusions}

This study presents a robust and scalable ensemble framework for the early diagnosis of AD based on MRI data. Our proposed method integrates three key techniques: transfer learning, weighted averaging, and stacked generalization. Together, these components form a powerful diagnostic pipeline that significantly outperforms conventional ML approaches and standalone convolutional neural networks. The experimental results presented in Tables~\ref{tab:ad_mci}, \ref{tab:mci_nc}, and \ref{tab:comparison} highlight the effectiveness of our model across various classification tasks involving AD, MCI, and NC subjects.

At the core of our approach is the utilization of diverse pretrained CNN architectures—ResNet50, NASNet, and MobileNet—which were fine-tuned end-to-end on brain MRI slices. These models capture different representational features of brain structure, providing the ensemble with both depth and breadth in learned features. The weighted averaging strategy ensures that more reliable base classifiers are emphasized during inference, while the stacked generalization mechanism learns higher-level patterns by combining base model predictions through a meta-learner. This two-level architecture enables the model to make more nuanced decisions, particularly in difficult cases such as distinguishing between MCI and NC.

Our improved ensemble strategy demonstrated superior performance across all evaluated tasks. Specifically, the model achieved 99.21\% accuracy in classifying AD vs. MCI and 91.02\% in MCI vs. NC—two of the most clinically significant tasks. Notably, the sensitivity for detecting early-stage conditions such as MCI was markedly improved, which is crucial for timely medical intervention. The area under the receiver operating characteristic curve remained consistently high (0.98--1.00), confirming the model's reliability across different operating thresholds.

Beyond raw performance, our methodology offers practical advantages for real-world clinical deployment. By leveraging transfer learning, the model requires less training data and computational resources while still achieving high accuracy. The ensemble structure provides robustness against overfitting and variability in imaging protocols. Additionally, our framework is modular and extensible—new base models or meta-learners can be added to the ensemble without altering the entire system. Despite its strengths, the proposed method is not without limitations. While the ADNI dataset provides a standardized benchmark, generalization to other clinical datasets has yet to be validated. MRI data is known to exhibit inter-center variability due to differences in scanner hardware, imaging protocols, and patient demographics. To address this challenge, future work will explore domain adaptation techniques to make the model robust across institutions. Furthermore, integrating federated learning can help in building generalized models without compromising data privacy, especially when training across multiple hospitals or countries. In conclusion, our ensemble methodology presents a significant advancement in computer-aided diagnosis for neurodegenerative diseases. It combines model diversity with intelligent fusion strategies to deliver a system that is not only accurate and stable, but also adaptable for clinical use. This work lays the foundation for future extensions that can include multi-modal data, interpretable AI components, and real-world deployment in early screening programs for Alzheimer’s Disease.

\subsection*{Acknowledgments}
We appreciate the service of ADNI for data support, and Arizona State University for partial funding and lab support.

\subsection*{Code Availability}
Code and supplementary materials are available at \url{https://github.com/FahadMostafa91/Hybrid_Deep_Ensemble_Learning_AD}.

\subsection*{Declaration of Interests}

The authors have no conflict of interest to report.

\subsection*{Ethics Approval}

There is no ethical approval needed due to the use of simulated and publicly available data.

\subsection*{Funding Statement}

The authors do not have funding to report.

\subsection*{Clinical Trial Registration}

The authors did not use clinical trial data directly. The authors used publicly available data with proper references in the text.

\subsection*{Informed Consent Statement}
This research was conducted on human subject data. Data were obtained from open sources.

\subsection*{Data Availability Statement} MRI data used in my research is publicly available from the Alzheimer’s Disease Neuroimaging Initiative
(ADNI) database (accessed on December 11th 2024).

\subsection*{Generative AI statement}

The author(s) declare that no Generative AI was used in the creation of this manuscript.

\bibliographystyle{unsrt}
\bibliography{bibliography}

\end{document}